\documentclass[aps,prd,preprint,groupedaddress,showpacs,showkeys,fleqn,11pt]{revtex4-1}
\usepackage{graphicx}
\usepackage{amsmath}
\usepackage{array}
\usepackage{longtable}
\usepackage{natbib}
\begin{document}
\title{Self-Gravitating Relativistic Models of Fermions\\ with Anisotropy and Cutoff Energy in their Distribution Function}
\author{Marco Merafina}
\email{marco.merafina@roma1.infn.it}
\affiliation{Department of Physics, University of Rome La Sapienza, Piazzale Aldo Moro 2, I-00185 Rome, Italy}
\author{Giuseppe Alberti}
\email{giuseppe.alberti@irsamc.ups-tlse.fr}
\affiliation{Laboratoire de Physique Th\'{e}orique (UMR 5152), IRSAMC, Universit\'{e} Paul Sabatier, 118 Route de Narbonne, F-31062 Toulouse Cedex 9, France}

\begin{abstract}
In this paper we study the equilibrium configurations of anisotropic self-gravitating fermions, by extending to general relativity the solutions obtained in a previous paper. This treatment also generalizes to anisotropic systems the relativistic self-gravitating Fermi gas model, by considering different degrees of anisotropy. We discuss some important characteristics of the models and the obtained density profiles, and generalize the relation between the anisotropy and the mass of particles in the relativistic regime. These relativistic models may also be applied to the study of superdense neutron stars with anisotropic pressure or super-Chandrasekhar white dwarfs generated by the presence of a magnetic field.
\end{abstract}

\pacs{47.75.+f -- 97.60.Jd -- 98.62.Gq}
\keywords{self-gravitating systems -- fermions -- anisotropy -- general relativity}
\maketitle

\section{INTRODUCTION} \label{introduction}

The theory of the general relativity (GR) is the rigorous way to describe properties and structure of systems kept bound by strong gravitational forces. Although a wide range of astrophysical objects can be analyzed only by using the classical mechanics (e.g., the globular clusters or anisotropic Newtonian systems; see \cite{bkmv09}), there are some cases in which it is necessary to use Einstein's field theory (e.g., the compact objects). However, a lot of phenomena do not require the full employement of the GR. In fact, it is sufficient to use an approximate method and consider two possibilities: the 1PN (first post-Newtonian) and the weak field approximations \cite{1972weinberg_book}. The 1PN approximation, where some examples of this application are in Refs. \cite{2011ago_ped_rc, ramos_2012}, gives corrections up to order $v^{2}/c^{2}$ (with $v$ the typical velocity in the system being considered and $c$ the speed of light) working for non relativistic particles ($v\ll c$); the weak field approximation, where examples can be found in Refs. \cite{ch_e_zhuk_2012, lu_huang2013}, is instead related to problems considering the gravitational radiation.

Other possibilities, in order to analyze the properties of self-gravitating systems in GR, are represented by solving the relativistic versions of the collisional and noncollisional Boltzmann equation or by considering a statistical approach. Important results of the first method are represented by the relativistic stellar clusters (see \cite{bkmv10}, hereafter BKMV10); for the second approach we have to mention the important papers of Fowler \cite{fowler26}, Chandrasekhar \cite{chandra31} and Oppenheimer and Volkoff \cite{opp_volk39}, that represent the first applications of the self-gravitating Fermi gas model in astrophysics.

More recent applications of the Fermi gas model in the framework of the GR have been proposed by Bili{\'c} and Viollier \cite{bil_v97, bil_v99}, who studied the general relativistic version of the Thomas-Fermi model and applied it to galactic dark halos, by supposing the existence of fermions spheres made by massive neutrinos ($m\sim 15$ keV). Nakajima and Morikawa \cite{nak_mor05} considered the equilibrium configurations of weakly interacting fully degenerate fermionic dark matter at various scales in the Universe, also finding a limiting mass in the range $2-30$ eV. Furthermore, Narain \emph{et al.} \cite{narain_etal_2006} proposed different models of compact stars, constituted by fermionic dark matter, finding the typical values for the masses of these stars by considering all the fermionic candidates for dark matter, from the heaviest to the lightest ones.

More refined models can be obtained by considering the presence of anisotropies in the distribution function characterizing the system under investigation (for an extended review of the effects of the local anisotropy in GR see Ref.\cite{herrera_sant97}). The sources of the anisotropy can be very different and are essentially connected to the presence of tensors that describe the pressure of a fluid or the effects of an external force on the system under consideration. One of the first important sources of the anisotropy is represented by the rotations: interesting consequences about the final phases of the evolutionary path of stars can be seen in Refs. \cite{pgit03a, pgit03b, tornap13}.

Because of the geometry of the $\pi$ modes, Sawyer and Scalapino \cite{sawscap73} argued the possibility that, in order to study the pion condensed phase configurations in the superdense nuclear matter, anisotropic distributions of pressure in the neutron stars could be included. Moved from this kind of motivations, Bowers and Liang \cite{bow_liang74} and Heintzmann and Hillebrandt \cite{hh75} first proposed anisotropic models for neutron stars (examples of more recent works on the same topic are given in Refs. \cite{dg02, harko_mak02, makharko03}).

Moved from the same reasons, but working in a different context, Bisnovatyi-Kogan \& Zel'dovich \cite{bkz69a, bkz69b} advanced analytic self-similar solutions for relativistic stellar clusters in presence of anisotropy: nevertheless, the formal solutions they found presented infinite central densities and infinite radii. Extended polytropic models for anisotropic systems can be found in Refs. \cite{herr_barreto13b, nguyen_ling13}, whereas examples of models for anisotropic general relativistic fluids are in Refs. \cite{hernunez_02, hernunez_04, herrera_etal_2004, nguyen_ped13}. Furthermore, some proposals connected to the galactic halos and the gravitational lensing of the dark matter can be found in Refs. \cite{boh_harko07, gallom_12}.

The presence of intense magnetic fields in the stars can have dramatic effects for what concerns the stability and the evolution of these objects: in fact, the possibility to have a ``super-Chandrasekhar white dwarf" \cite{2006howell_al} (that could explain the overluminous type Ia supernovae \cite{2010scalzo_et, 2007hicken_et, 2011silverman_et, 2011taubenberger_et, 2009yamanaket}), can be explained in this way. The presence of the magnetic field could lead the star beyond the well known Chandrasekhar limiting mass ($M_{Ch}=5.75 \mu_e^{-2} M_{\odot}$, \cite{chtoop64}) and make it as the progenitor of these overluminous supernovae. In these situations the magnetic field may be treated as an anisotropic fluid.

Although there are the first theoretical explanations of this puzzling behavior of the white dwarfs in presence of anisotropies \cite{dasmukh2013a, dasmukh2013b, herrera_etal_2014}, we argue the possibility that a description of these objects in terms of fermions (as it is known, white dwarfs stars - like neutron stars - are formed by degenerate fermionic matter) is more precise and rigorous. For these reasons in this work we extend the Newtonian models of the collisionless semidegenerate Fermi gas, described in our previous paper (\cite{paperI}, hereafter Paper I), to GR. In Sec. \ref{model} we introduce the distribution function and define the thermodynamic quantities, as the tensor pressure and the density, to solve the equilibrium equations. In Sec. \ref{results} we present the results of the numerical integration, by discussing the characteristics of the models studied. In Sec. \ref{limits} we derive a relation (valid in the limit of full degeneracy) between the mass of the particles and the anisotropy in the distribution function. Finally, in Sec. \ref{conclusions}, we draw some conclusions.

\section{THE MODEL} \label{model}
\subsection{Distribution Function and Useful Variables} \label{submodel_1}
The distribution function has the form (see Paper I)
\begin{equation} \label{f_m_a}
\begin{split}
f & = \frac{g}{h^{3}}\,\left(1+\frac{L^{2}}{L_{c}^{2}}\right)^l\,\,\frac{1 - e^{(\epsilon - \epsilon_{c})/kT_r}}{e^{(\epsilon - \mu)/kT_r} + 1} & \text{ for } \epsilon \leq \epsilon_{c} \ , \\
f & = 0 & \text{for } \epsilon > \epsilon_{c} \ ,
\end{split}
\end{equation}
where $T_r$, in relativistic regime, is the local temperature \cite{1934tolman_book}, $g=2s+1$ is the spin multiplicity of quantum states, $h$ and $k$ are the Planck and Boltzmann constants, respectively. $L_c = mcr_a$ is a constant depending on the anisotropy radius $r_{a}$ ($c$ is the speed of light), defined as the radius beyond which the effects of the anistropy become relevant: in the limit $r_a\rightarrow\infty$ the isotropy is completely recovered, as to mean that it is necessary an infinite distance to reach the anisotropy (for more details see \cite{michie61,michie63}); $L = mv_t r=p_t r$ is the angular momentum of a single particle ($m$, $r$, and $v_{t}$ are, respectively, the particle mass, the radial coordinate, and the tangential velocity); $\mu$ is the chemical potential, defined by the well known relation $\mu=(\partial U/\partial N)_{S,V}$; $\epsilon$ is the kinetic energy, while $\epsilon_{c}$ is the cutoff kinetic energy, i.e. the maximal kinetic energy that a particle can have at a given radius $r$.

The presence of $L$ in the distribution function does not necessary require the relaxation of the spatial spherical symmetry: in fact $L$ is connected to the momentum phase space, and, in only this space, the anisotropy plays a role. It is thus possible to have an anisotropic system in which the spatial distribution of the matter is spherically symmetric and for this reason we use the Schwarzschild metric
\begin{equation} \label{schwarzschild_metric}
ds^{2} = e^{\nu}c^{2}dt^{2} - e^{\lambda}dr^{2} - r^{2}(d\psi^{2} + sin^{2}\psi\,d\varphi^{2})\ ,
\end{equation}
and the equations of the gravitational equilibrium are given by \cite{bkz69a}
\begin{equation} \label{eq_bkz}
\begin{split}
& \frac{dP_{rr}}{dr} = -\frac{G}{c^2} \frac{(P_{rr} + \rho c^{2})(M_{r} c^{2} + 4\pi P_{rr} r^{3})}{r\,(rc^2-2GM_r)} - \frac{2}{r} (P_{rr}-P_{t})
\ ,\\ & \frac{dM_r}{dr} = 4\pi\rho r^{2} \ ,
\end{split}
\end{equation}
with the conditions $P_{rr}(0)$ = $P_{rr0}$ and $M_{r}(0) = 0$. Here $G$ is the gravitational constant, $P_{rr}$ and $P_t$ are, respectively, the radial and tangential components of the pressure tensor, $\rho c^2$ the total energy density and $M_r$ the mass within a given radius $r$. The metric coefficients of Eq.(\ref{schwarzschild_metric}) are defined by the following expressions
\begin{equation} \label{eq_lambda}
e^{\lambda}=\left(1-\frac{2GM_r}{rc^2}\right)^{-1} \,,
\end{equation}
\begin{equation} \label{eq_nu}
\frac{d\nu}{dr} =\frac{2G}{c^2}\,\frac{M_r c^2 +4\pi P_{rr}r^3}{r\,(rc^2 - 2GM_r)}\,,
\end{equation}
\begin{equation} \label{lambda_nu_bordo}
e^{\nu_{R}} = e^{-\lambda_{R}} = 1 - \frac{2GM}{Rc^{2}} \,.
\end{equation}

To solve the Eqs.(\ref{eq_bkz}) and evaluate the thermodynamic functions, let us introduce the variables
\begin{equation} \label{variables}
\begin{split}
\epsilon & = \sqrt{p^{2}c^{2} + m^{2}c^{4}} - mc^{2} \,, \quad \epsilon_{c} = \sqrt{p_{c}^{2}c^{2} + m^{2}c^{4}} - mc^{2} \,, \quad T = T_r e^{\nu/2} \, ,\\
x & = \frac{\epsilon}{kT_{r}} \,,  \quad \quad y = \frac{\epsilon}{mc^{2}} \,, \quad \quad W = \frac{\epsilon_c}{kT_r} \,, \quad \quad \theta = \frac{\mu}{kT_{r}} \,, \quad \quad \beta = \frac{kT_{R}}{mc^{2}} \, .
\end{split}
\end{equation}
Here, $\epsilon$ and $\epsilon_{c}$ are the same variables appearing in the distribution function, whereas $T$ is the temperature of the system ``measured by an infinitely-remote observer", constant all over the equilibrium configuration \cite{1934tolman_book}; $x$, $y$, $W$, $\theta$ and $\beta$ are dimensionless variables. In particular, $\theta$ and $\beta$ are, respectively, the degeneracy and the relativistic temperature parameters. Moreover, on the basis of the energy conservation, we define
\begin{equation} \label{conservation}
\begin{split}
& E\equiv(\epsilon + mc^{2}) e^{\nu/2} = \mbox{constant}\,, \\
& E_c\equiv(\epsilon_{c} + mc^{2}) e^{\nu/2} = mc^{2} e^{\nu_{R}/2} \,, \\
& \Upsilon\equiv(\mu + mc^{2}) e^{\nu/2} = (\mu_{R} + mc^{2}) e^{\nu_{R}/2} \,.
\end{split}
\end{equation}
Using relations (\ref{variables}) and (\ref{conservation}) we obtain also
\begin{equation} \label{eq_kT_beta}
\frac{mc^{2}}{kT_{r}} = \frac{1 - \beta W}{\beta} \,,
\end{equation}
that we can rewrite as
\begin{equation} \label{eq_beta_nu}
1 - \beta W = \frac{\beta mc^{2}}{kT_{r}} = \frac{T_{R}}{T_{r}} = \frac{T e^{-\nu_{R}/2}}{T e^{-\nu/2}} = e^{\frac{\nu - \nu_{R}}{2}} \,.
\end{equation}
From Eq.(\ref{eq_beta_nu}) we have the constraint $0\leq\beta W<1$ (see Refs. \cite{meruf89, meruf90}) and moreover, we get
\begin{equation} \label{eq_beta_nu2}
e^{\nu} = e^{\nu_{R}} {(1 - \beta W)}^{2} \quad \quad \mbox{and} \quad \quad \frac{d\nu}{dr} = -\frac{2\beta}{1 - \beta W} \frac{dW}{dr} \,.
\end{equation}
Substituting Eq.(\ref{eq_nu}) into Eq.(\ref{eq_beta_nu2}) we obtain, instead of the first of the Eqs.(\ref{eq_bkz}),
\begin{equation} \label{eq_TOV_W}
\frac{dW}{dr} = -\frac{G}{c^2}\,\left(\frac{1-\beta W}{\beta}\right)\, \frac{M_r c^2 +4\pi P_{rr}r^3}{r\,(rc^2-2GM_r)}\, ,
\end{equation}
with $W(0)=W_0$\,.

\subsection{Thermodynamic Quantities and Gravitational Equilibrium} \label{submodel_2}
The thermodynamic variables are defined by relations (see BKMV10)
\begin{equation} \label{n_1}
n = \frac{2 \pi g}{h^{3}}\sum_{k=0}^{l} \binom{l}{k} \left(\frac{r}{L_c}\right)^{2k} \int_0^{\pi} (\sin\psi)^{2k+1} d\psi \int_0^{p_{c}} \frac{1 - e^{(\epsilon - \epsilon_{c})/kT_r}}{e^{(\epsilon -\mu)/kT_r}+1}\,p^{2k+2}dp \,,
\end{equation}
\begin{equation} \label{rho_1}
\rho c^{2} = \frac{2 \pi g}{h^{3}} \sum_{k=0}^{l} \binom{l}{k} \left(\frac{r}{L_c}\right)^{2k}\int_0^{\pi} (\sin\psi)^{2k+1} d\psi \int_0^{p_{c}}\frac{1-e^{(\epsilon -\epsilon_c)/kT_r}}{e^{(\epsilon -\mu)/kT_r}+1}\,p^{2k+2}\sqrt{p^2 c^2 + m^2 c^4}\,dp \,,
\end{equation}
\begin{equation} \label{Prr_1}
P_{rr} = \frac{2 \pi gc^2}{h^{3}} \sum_{k=0}^{l} \binom{l}{k} \left(\frac{r}{L_c}\right)^{2k}\int_0^{\pi}(\sin\psi)^{2k+1}(\cos\psi)^2 d\psi \int_0^{p_{c}} \frac{1-e^{(\epsilon -\epsilon_c)/kT_r}}{e^{(\epsilon -\mu)/kT_r}+1}\,\frac{p^{2k+4}}{\sqrt{p^2 c^2 +m^2 c^4}}\,dp \,,
\end{equation}
\begin{equation} \label{Pt_1}
P_{t} = \frac{\pi gc^2}{h^{3}} \sum_{k=0}^{l} \binom{l}{k} \left(\frac{r}{L_c}\right)^{2k} \int_0^{\pi} (\sin\psi)^{2k+3} d\psi \int_0^{p_{c}} \frac{1-e^{(\epsilon -\epsilon_c)/kT_r}}{e^{(\epsilon -\mu)/kT_r}+1}\, \frac{p^{2k+4}}{\sqrt{p^2 c^2 +m^2 c^4}}\,dp \,,
\end{equation}
where $n$ represents the number density of the particles while $\rho c^2$, $P_{rr}$ and $P_t$ have been previously defined. In the Eqs.(\ref{n_1})-(\ref{Pt_1}) we have used the relations $p_r=p\cos\psi$ and $p_t=p\sin\psi$ for the components of the momentum $p$, whereas $p_c$ is the value of the momentum corresponding to the cutoff energy, and we have rewritten the part of the distribution function depending on the angular momentum by using the Newton binomial relation
\begin{equation} \label{newton}
\left(1 + \frac{L^{2}}{L_{c}^{2}}\right)^{l} = \sum_{k=0}^{l} \binom{l}{k} \left(\frac{L}{L_{c}}\right)^{2k} , \quad \quad \mbox{with} \quad \quad \binom{l}{k} = \frac{l!}{k!(l-k)!}, \quad 0! = 1 \,.
\end{equation}

To transform the integrals of Eqs.(\ref{n_1})-(\ref{Pt_1}) into a more suitable form, following BKMV10, it is more convenient to use the variables $x$ and $y$ [see Eqs.(\ref{variables})] instead of $p/mc$
\begin{equation} \label{p_x}
\frac{\sqrt{p^2 c^2 +m^2 c^4}}{mc^2}=\frac{\epsilon}{mc^2}+1=y+1,
\quad\quad \mbox{where} \quad\quad \frac{p}{mc}=\sqrt{y\,(y+2)} \,.
\end{equation}
By differentiating and using Eqs.(\ref{variables}) we obtain
\begin{equation} \label{p_x2}
\frac{dp}{mc}=\frac{y+1}{\sqrt{y\,(y+2)}}\,dy=\left(\frac{\beta /2}
{1-\beta W}\right)^{1/2}\left(1+\frac{\beta x}{1-\beta W}\right)
\left(1+\frac{\beta x/2}{1-\beta W}\right)^{-1/2}\frac{dx}{\sqrt x}
\end{equation}
and, substituting this result into the expressions of the thermodynamic functions, we get
\begin{equation} \label{n_2}
n=\frac{\pi g m^{3}c^{3}}{h^{3}}\sum_{k=0}^{l}\binom{l}{k} \left(\frac{r}{r_a}\right)^{2k} A_{k} \left(\frac{2\beta}{1-\beta W}\right)^{k+\frac{3}{2}} I_{nk} \,,
\end{equation}
\begin{equation} \label{rho_2}
\rho c^2 =\frac{\pi gm^{4}c^{5}}{h^{3}}\sum_{k=0}^{l} \binom{l}{k} \left(\frac{r}{r_a}\right)^{2k} A_{k}\left(\frac{2\beta}{1-\beta W}
\right)^{k+\frac{3}{2}}I_{\rho k} \,,
\end{equation}
\begin{equation} \label{Prr_2}
P_{rr}=\frac{\pi gm^{4}c^{5}}{h^{3}}\sum_{k=0}^{l} \binom{l}{k} \left(\frac{r}{r_a}\right)^{2k} \left(A_{k} - A_{k+1}\right) \left(\frac{2\beta}{1-\beta W}\right)^{k+\frac{5}{2}}I_{Pk}\,,
\end{equation}
\begin{equation} \label{Pt_2}
P_{t}=\frac{\pi gm^{4}c^{5}}{2h^{3}}\sum_{k=0}^{l} \binom{l}{k} \left(\frac{r}{r_a}\right)^{2k} A_{k+1}\left(\frac{2\beta}{1-\beta W}\right)^{k+\frac{5}{2}}I_{Pk}\,.
\end{equation}
Here, the $A_{k}$ coefficients \cite{1985bs_book} and the integrals $I_{nk}$, $I_{\rho k}$ and $I_{Pk}$ are defined, respectively, by
\begin{equation} \label{ak}
A_k =\int_0^\pi (\sin\psi)^{2k+1} d\psi = 2\sum_{i=0}^{k}\binom{k}{i} \frac{(-1)^i}{2i+1}\,,
\end{equation}
\begin{equation} \label{I_nk}
I_{nk}=\int_0^W g(x,W)\left(1+\frac{\beta x}{1-\beta W}\right)\left(1+\frac{\beta x/2}{1-\beta W}\right)^{k+\frac{1}{2}}x^{k+\frac{1}{2}}dx\,,
\end{equation}
\begin{equation} \label{I_rhok}
I_{\rho k}=\int_0^W g(x,W)\left(1+\frac{\beta x}{1-\beta W}\right)^2 
\left(1+\frac{\beta x/2}{1-\beta W}\right)^{k+\frac{1}{2}}x^{k+\frac{1}{2}} dx\,,
\end{equation}
\begin{equation} \label{I_Pk}
I_{Pk}=\int_0^W g(x,W)\left(1+\frac{\beta x/2}{1-\beta W}\right)^{k+\frac{3}{2}} x^{k+\frac{3}{2}}dx\,,
\end{equation}
and the $g(x,W)$ function is given by
\begin{equation} \label{g_x_W}
g(x,W)=\frac{1-e^{x-W}}{e^{x-\theta}+1}=\frac{1-e^{x-W}}{e^{x-W-\theta_R}+1}\,,
\end{equation}
where $\theta=W+\theta_R$ remains valid also in relativistic regime and $\theta_R=\theta(R)$ (see Paper I). In particular, the first three values of the $A_{k}$ coefficients $A_{0}$, $A_{1}$, and $A_{2}$ are 2, 4/3, and 16/15, respectively.

\subsection{Dimensionless Variables} \label{submodel_3}
Following the same procedure of the Newtonian case, let us introduce the dimensionless variables
\begin{equation} \label{dimensionless}
r = \xi \tilde{r}\,,\quad r_a =\xi a\,,\quad n=\frac{c^2 \tilde{n}}{Gm\xi^{2}}\,,\quad\rho c^2=\frac{c^4\tilde{\rho}}{G\xi^{2}}\,,\quad P_{rr}= \frac{c^4 \tilde{P}_{rr}}{G\xi^{2}}\,,\quad P_{t}=\frac{c^4 \tilde{P}_{t}}
{G\xi^{2}}\,,\quad M_{r}= \frac{c^2\xi\tilde{M}_r}{G}\,,
\end{equation}
where $\xi = {(h^{3}/gcGm^{4})}^{1/2}$ and {\itshape a} is the dimensionless anisotropy radius, hereafter named anisotropy parameter. The thermodynamic quantities in dimensionless form are
\begin{equation} \label{n_less}
\tilde{n} = \pi\sum_{k=0}^{l}\binom{l}{k}\left(\frac{\tilde{r}}{a}\right)^{2k} A_{k} \left(\frac{ 2\beta}{1-\beta W}\right)^{k+\frac{3}{2}} I_{nk}\,,
\end{equation}
\begin{equation} \label{rho_less}
\tilde{\rho} =\pi\sum_{k=0}^{l}\binom{l}{k}\left(\frac{\tilde{r}}{a}\right)^{2k} A_{k}\left(\frac{ 2\beta}{1-\beta W}\right)^{k+\frac{3}{2}} I_{\rho k}\,,
\end{equation}
\begin{equation} \label{Prr_less}
\tilde{P}_{rr}=\pi\sum_{k=0}^{l}\binom{l}{k}\left(\frac{\tilde{r}}{a}\right)^{2k}\left(A_{k}-A_{k+1}\right)\left(\frac{2\beta}{1-\beta W}\right)^{k+\frac{5}{2}} I_{Pk}\,,
\end{equation}
\begin{equation} \label{Pt_less}
\tilde{P}_{t}=\frac{\pi}{2}\sum_{k=0}^{l}\binom{l}{k}\left(\frac{\tilde{r}}{a}\right)^{2k} A_{k+1}\left(\frac{2\beta}{1-\beta W}\right)^{k+\frac{5}{2}} I_{Pk}\,.
\end{equation}
The equilibrium equations then become
\begin{equation} \label{equilibrium_less}
\begin{split}
& \frac{dW}{d\tilde{r}}= -\left({\frac{1-\beta W}{\beta}}\right) \frac{\tilde{M}_r+4\pi\tilde{P}_{rr}\tilde{r}^3}{\tilde{r}\,(\tilde{r} -2\tilde{M}_r)}\,, \\
& \frac{d\tilde{M}_{r}}{d\tilde{r}} = 4\pi\tilde{\rho}\tilde{r}^2\,,
\end{split}
\end{equation}
with the initial conditions $W(0) = W_{0}$ and $\tilde{M}_{r}(0) = 0$.

\section{RESULTS OF THE NUMERICAL INTEGRATION} \label{results}
In this paper we calculate the equilibrium configurations by considering $l=1$ in the expression of the distribution function (\ref{f_m_a}).
\subsection{Evidences of the Anisotropy} \label{subresult_1}
To explicitly analyze the effects of the presence of the anisotropy in the equilibrium configurations, in analogy with the Newtonian treatment, it is useful to define the parameter $\eta$
\begin{equation} \label{eta}
\eta = \frac{2\langle v_r^2\rangle}{\langle v_t^2\rangle}=\frac{P_{rr}}{P_t}=\frac{\tilde{P}_{rr}}{\tilde{P}_t}= \frac{1+\frac{4}{5}\left(\frac{\beta}{1-\beta W}\right)\frac{I_{P1}}{I_{P0}}\left(\frac{\tilde{r}}{a}\right)^2}{1+\frac{8}{5}\left(\frac{\beta}{1-\beta W}\right)\frac{I_{P1}}{I_{P0}}\left(\frac{\tilde{r}}{a}\right)^2}\,.
\end{equation}
Since this definition is the same shown in BKMV10, we expect a similar trend, by analyzing the behavior of $\eta$ starting from the center towards the boundary of the equilibrium configurations. Figure \ref{fig:eta_beta}, in fact, confirms what we expect: both in the limit $r \rightarrow 0$ (at the center) and in the limit $r \rightarrow R$ (at the edge), we see that the ratio $P_{rr}/P_{t} \rightarrow 1$ and $\eta$ reaches its maximum value ($\eta_{\text{max}} = 1$), by showing a prevalence of isotropic motion of the particles. In the intermediate zones, we instead can note a decrease of $\eta$ until to its minimum value (which cannot be less than $\eta_{\text{min}} = 0.5$), clear gauge of the prevalence of tangential motion.

From Eq.(\ref{eta}) we can also study the behavior of $\eta$ as a function of the temperature parameter $\beta$, both in the limit $\beta \rightarrow 0$ and in the limit $\beta \rightarrow \infty$. In the first case, we note that $\beta/(1-\beta W) \rightarrow 0$ and $\eta \rightarrow 1$ (furthermore the expressions of $I_{P0}$ and $I_{P1}$ in Eq.(\ref{eta}) tend to the corresponding ones in the Newtonian limit). In the second case, we see that $\beta/(1-\beta W) \gg 1$ and thus $\eta \rightarrow 0.5$.


\subsection{Mass - Central Density Diagrams} \label{subresult_2}
In this section we aim at describing the mass - central density diagrams. In Fig. \ref{fig:M_rho1} we represented the effects of the anisotropy with changing the parameter $a$ for a fixed value of $\beta$. We can note that a larger degree of anisotropy requires the existence of configurations with smaller masses. Indeed, if we fix the value of $a$ and vary the value of $\beta$ (see Fig. \ref{fig:M_rho2}), we can note that configurations with high values of $\beta$ have a total mass generally larger than configurations with small ones (i.e., the Newtonian limit). In particular, it is interesting to note how the value of $\beta$ affects the level of degeneracy in the equilibrium configurations, due to the constraint (\ref{eq_beta_nu}) and the condition $\theta\leq W$ that implies $\theta_R\leq 0$ (for more details see Paper I).

In Fig. \ref{fig:M_rho3} we show the behavior of the equilibrium configurations in the isotropic limit, which is recovered for $a \rightarrow \infty$ (but $a = 1$ already constitutes an excellent approximation of this limit). In this diagram we have pointed our attention to the influence of the degeneracy level on the equilibrium configurations. To do this we have constructed some curves for $\beta = 10^{-5}$ and $\beta = 10^{3}$ choosing four values of $\theta_R$: $0, -2.31, -5, -10$. Considering the curves at $\beta = 10^{-5}$, in the degenerate limit and for large values of the central density, we observe the four curves follow the same behavior whereas, for smaller values of $\rho_0$, we note a split of them to correspond with the bifurcation point. In particular, the curve with $\theta_R = -10$ reaches higher values of the mass than the other three and present a local maximum before the bifurcation point. When we refer to the semidegenerate limit we do not see an overlapping of the curves and, like the previous case, the curve with $\theta_R = -10$ reaches again the highest values of the mass. For $\beta = 10^{3}$ we observe a more regular behavior where the curves at smaller values of $\theta_R$ show higher values of the mass.

\subsection{Density Profiles} \label{subresult_3}
\begin{figure}
\centering
\includegraphics[scale = 0.45]{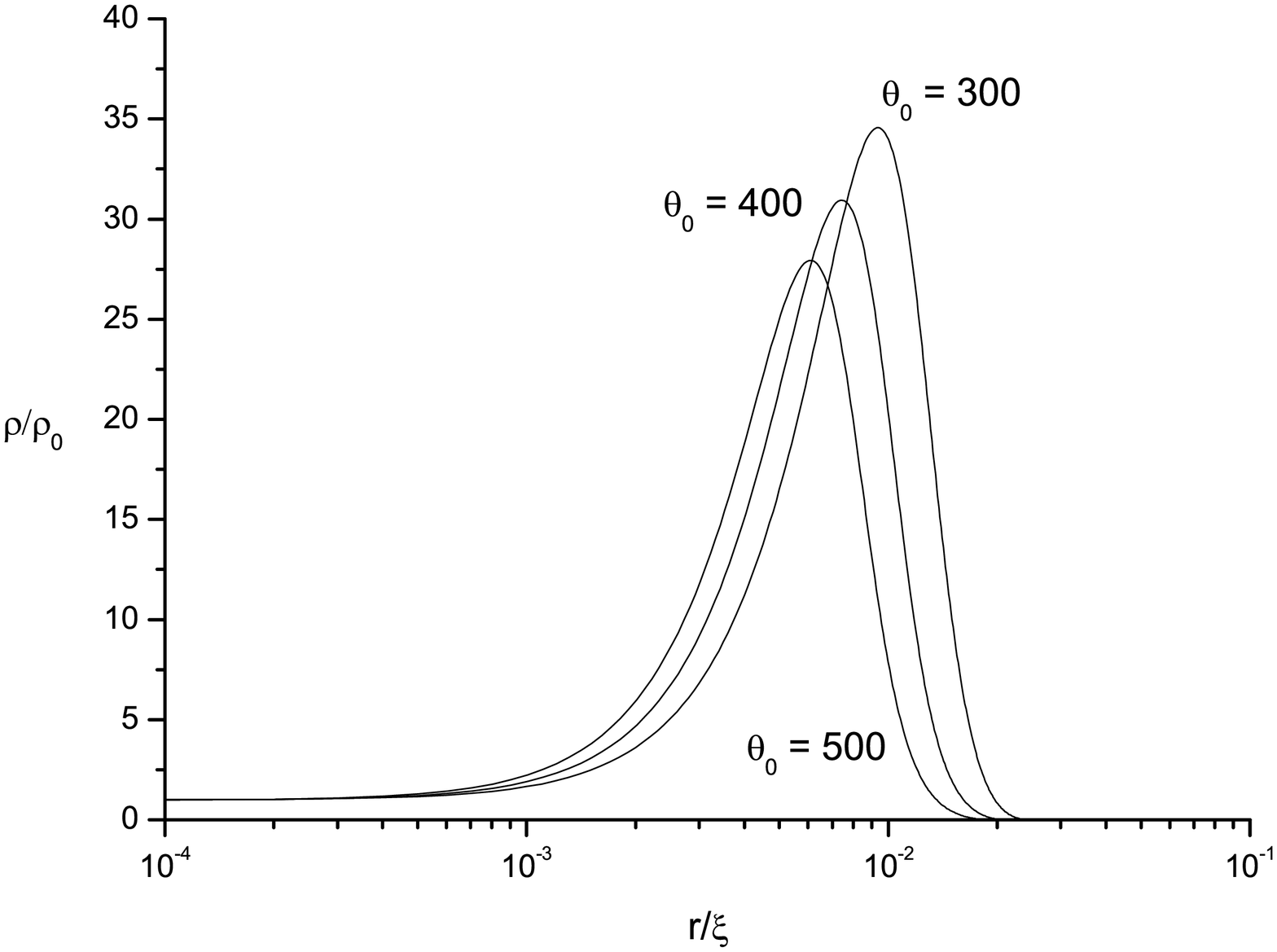}
\caption{Relative density $\rho/\rho_{0}$ as a function of the dimensionless radial coordinate $r/\xi$ at different values of $\theta_0$, for \emph{a} = $10^{-3}$, $\beta W_0=0.5$ and $\beta = 10^{-3}$.}
\label{fig:profile1}
\end{figure}

By integrating Eqs.(\ref{equilibrium_less}), we can construct the density profiles of the equilibrium configurations. Expressing in terms of dimensionless quantities and remembering that $l=1$, we have
\begin{equation} \label{rho_tilde}
\tilde{\rho}=4\sqrt{2}\pi\left(\frac{\beta}{1-\beta W}\right)^{3/2}
\left\{I_{\rho 0} +\frac{4}{3}\left(\frac{\beta}{1-\beta W}\right)
I_{\rho 1}\left(\frac{\tilde{r}}{a}\right)^2\right\}\,.
\end{equation}
The variation of the anisotropy parameter remarkably influences the behavior of the density function. From Eq.(\ref{rho_tilde}), in the limit $a \rightarrow 0$, it follows immediately that the second addend prevails, implying a general increase of the density $\tilde{\rho}$ and, in particular, of its maximum value. In Figs. \ref{fig:profile1}, \ref{fig:profile2} and \ref{fig:profile3} we have represented the quantity $\rho/\rho_{0}$ as a function of the dimensionless radial coordinate $r/\xi$, for different values of \emph{a}, $\beta$, $W_{0}$ and $\theta_{0}$.

In Fig. \ref{fig:profile1} we show the behavior of the density profiles, once fixed the values of $a$ and $\beta$ by varying the value of the central degeneracy parameter $\theta_0$. One can see how the value of the maximum increases by decreasing $\theta_0$ and its position tends to move in the direction of the periphery of the configuration. The trend of the density profiles, also in relativistic regime, shows the existence of hollow configurations and confirm the results obtained by Nguyen and Pedraza \cite{nguyen_ped13} and by Ralston and Smith \cite{1991ralsmith}, giving a clear indication that the presence of the anisotropy is the reason of this kind of configurations.

Figs. \ref{fig:profile2} and \ref{fig:profile3} show the influence of the temperature parameter $\beta$ on the density profiles. If we look at Eq.(\ref{rho_tilde}) we see that, in the limit $\beta \rightarrow 0$, the second term in the sum becomes negligible and we recover the typical behavior of the isotropic systems. On the contrary, in the limit $\beta \rightarrow \infty$, the curve comes back to the behavior of the hollow systems. In Tables \ref{tab:table1}, \ref{tab:table2} and \ref{tab:table3} we summarize some results of the numerical integration for particular values of $a$, $\beta W_0$, $\beta$ and $W_0$, starting from the equilibrium configurations where the general relativistic effects (according to the value of $\beta W_0$) and the anisotropy are not very significant, up to configurations where the degree of relativity and anisotropy is very high.

\section{LIMITS ON THE PARTICLE MASS} \label{limits}
In Paper I, we derived, in the limit of full degeneracy, an analytical expression relating the mass of particles with the anisotropy in the momentum distribution. According to the relation obtained, an increase of the anisotropy within the distribution of velocities provoked a decrease of the lower limit of the particle mass. On the contrary, when the system recovered the isotropy, we noted an increase of the limiting mass. To extend this result to the GR regime, we have to consider a new parameter not considered in the Newtonian regime, i.e. the temperature parameter $\beta$. Let us rewrite the definition of the density $\rho$ as
\begin{equation} \label{rho_bis}
\rho=\frac{\pi gm^4 c^3}{h^3}\sum_{k=0}^{l}\binom{l}{k}\left(\frac{r}{r_a}\right)^{2k} A_{k}\left(\frac{2\beta}{1-\beta W}\right)^{\frac{2k+3}{2}}I_{\rho k}\,.
\end{equation}
In the limit of full degeneracy $g(x,W)\rightarrow 1$ and, using Eq.(\ref{rho_bis}), we can write
\begin{equation} \label{rho_deg}
\rho\leq\frac{2\pi gm^4 c^3\beta^{3/2}}{\alpha^4 h^3}\sum_{k=0}^{l} \binom{l}{k}\left(\frac{r}{r_a}\right)^{2k} A_{k}\left(\frac{\beta}{\alpha^2}\right)^{k}\int_0^W\left(\beta x+\alpha\right)^2\left[x
\left(\beta x+2\alpha\right)\right]^{\frac{2k+1}{2}}dx\,,
\end{equation}
where $\alpha=1-\beta W$. Making the substitution $z=\beta x+\alpha$, the integral in Eq.(\ref{rho_deg}) can be written as
\begin{equation}\label{J_rhok1}
\int_0^W\left(\beta x+\alpha\right)^2\left[x\left(\beta x+2\alpha\right)\right]^{\frac{2k+1}{2}}dx =\frac{\beta^{-\frac{2k+3}{2}}}{(2k+3)}\left[\left(1-\alpha^2\right)^{\frac{2k+3}{2}}-\int_\alpha^1\left(z^2 -\alpha^2\right)^{\frac{2k+3}{2}}dz\right]\,,
\end{equation}
where, with the substitution $z=\alpha\cosh y$,
\begin{equation}\label{J_rhok2}
\begin{split}
&\int_\alpha^1\left(z^2 -\alpha^2\right)^{\frac{2k+3}{2}}dz=\frac{1}{2k+4}
\quad\times \\
&\left\{\left(1-\alpha^2\right)^{\frac{2k+3}{2}}-\frac{2k+3}{2k+2}\left[\alpha^2\left(1-\alpha^2\right)^{\frac{2k+1}{2}}-(2k+1)\alpha^{2k+4}\int_0^{\ln\left(\frac{1+\sqrt{1-\alpha^2}}{\alpha}\right)}(\sinh y)^{2k}dy\right]\right\}\,.
\end{split}
\end{equation}
Thus, the integral is transformed into
\begin{equation}\label{J_rhok3}
\begin{split}
&\int_0^W \left(\beta x+\alpha\right)^{2}\left[x\left(\beta x+2\alpha\right)\right]^{\frac{2k+1}{2}}dx=\frac{\beta^{-\frac{2k+3}{2}}}{4(k+1)
(k+2)}\quad\times \\
&\left\{\left(1-\alpha^2\right)^{\frac{2k+1}{2}}\left[2(k+1)-
(2k+1)\alpha^2\right]-(2k+1)\alpha^{2k+4}\int_0^{\ln\left(\frac{1+\sqrt{1-\alpha^2}}{\alpha}\right)}(\sinh y)^{2k}dy\right\}
\end{split}
\end{equation}
and Eq.(\ref{rho_deg}) becomes
\begin{equation} \label{rho_deg2}
\begin{split}
& \rho\leq \frac{\pi gm^{4}c^{3}}{2h^{3}} \sum_{k=0}^{l}\binom{l}{k} \left(\frac{r}{r_{a}}\right)^{2k}\frac{A_k}{\alpha^{2k+4}(k+1)(k+2)}\quad \times \\
&\left\{\left(1-\alpha^2\right)^{\frac{2k+1}{2}}\left[2(k+1)-(2k+1)\alpha^{2}\right]-(2k+1)\alpha^{2k+4}\int_0^{\ln\left(\frac{1+\sqrt{1-\alpha^2}}{\alpha}\right)}(\sinh y)^{2k}dy\right\}\,.
\end{split}
\end{equation}
Equation (\ref{rho_deg2}) is a general expression corresponding to any value of $l$. In our case $l=1$, then we obtain
\begin{equation} \label{rho_deg3}
\rho\leq\frac{\pi gm^4 c^3}{2\alpha^4 h^3}\left\{\sqrt{1-\alpha^2}\left[2-\alpha^2 +\left(\frac{r^2}{r_a^2}\right)\frac{8-14\alpha^2 +3\alpha^4}{9 \alpha^2}\right] + \frac{\alpha^4}{3}\left(\frac{r^2}{r_a^2}-3\right)\ln\left(\frac{1+\sqrt{1-\alpha^2}}{\alpha}\right)\right\}
\end{equation}
and, solving for the mass, we have
\begin{equation} \label{massa_1}
\begin{split}
& m\geq\left(\frac{2\rho h^3}{\pi gc^3}\right)^{1/4}\alpha
\quad\times \\
&\left\{\sqrt{1-\alpha^2}\left[2-\alpha^2 +\left(\frac{r^2}{r_a^2}\right)\frac{8-14\alpha^2 +3\alpha^4}{9\alpha^2}\right]+\frac{\alpha^4}{3}\left(\frac{r^2}{r_a^2}-3\right)\ln\left(\frac{1+\sqrt{1-\alpha^2}}{\alpha}\right)\right\}^{-1/4}\,.
\end{split}
\end{equation}
We can find the expression from Eq.(\ref{massa_1}) corresponding to the center of the equilibrium configuration where $r=0$ and $\alpha_0 =1-\beta W_0$. We get
\begin{equation} \label{massa_2}
m\geq\left(\frac{2\rho_0 h^3}{\pi gc^3}\right)^{1/4}\alpha_0\left[
(2-\alpha_0^2)\sqrt{1-\alpha_0^2}-\alpha_0^4\,\ln\left(\frac{1+\sqrt{1-\alpha_0^2}}{\alpha_0}\right)\right]^{-1/4}\,,
\end{equation}
even if this expression (that for $\rho_0\sim 10^{15}$ g/cm$^{3}$, $g=2$, $\beta W_0 =0.117$ gives a particle mass $m\geq 1.68\times 10^{-24}$ g $\approx 939.5$ MeV) does not take into account the effects of the anisotropy which are not exhibited at the center of the equilibrium configurations. In order to have an evidence of these effects, following a similar approach as in Paper I, we get the expression from the Eq.(\ref{massa_1}) corresponding to the half mass radius $r=r_h$, where $\alpha_h =1-\beta W_h$ is the value for $r=r_h$ and $\rho=\rho(r_h)\equiv\rho_h$. Then
\begin{equation} \label{massa_3}
\begin{split}
& m\geq\left(\frac{2\rho_h h^3}{\pi gc^3}\right)^{1/4}\alpha_h\quad
\times \\
&\left\{\sqrt{1-\alpha_h^2}\left[2-\alpha_h^2+\left(\frac{r_h^2}{r_a^2}\right)\frac{8-14\alpha_h^2 + 3\alpha_h^4}{9\alpha_h^2}\right]+\frac{\alpha_h^4}{3}\left(\frac{r_h^2}{r_a^2}-3\right)\ln\left(\frac{1+\sqrt{1-\alpha_h^2}}{\alpha_h}\right)\right\}^{-1/4}\,.
\end{split}
\end{equation}
In this case, for obtaining the same condition on the mass $m\geq 1.68\times 10^{-24}$ g, we must have for example $\rho_0\sim 2.94\times 10^{15}$ g/cm$^{3}$, $\beta W_0 = 0.5$, $a=10^{-5}$ and $\beta W_h=0.364$ (or $\alpha_h=0.636$). We may also rewrite Eq.(\ref{massa_1}) in the form $m \geq m_*\, F(\alpha_0,a,\tilde r)$ by defining a ``dimensional" mass
\begin{equation} \label{massa_4}
m_* = \left(\frac{2\rho_0 h^3}{\pi gc^3}\right)^{1/4}\,,
\end{equation}
where the dimensionless function $F(\alpha_0,a,\tilde r)$ is given by
\begin{equation} \label{F_abeta}
\begin{split}
&F(\alpha_0,a,\tilde r)=\left(\frac{\tilde\rho}{\tilde\rho_0}\right)^{1/4}\alpha\quad\times \\
&\left\{\sqrt{1-\alpha^2}\left[2-\alpha^2 +\left(\frac{{\tilde r}^2}{a^2}\right)\frac{8-14\alpha^2 +3\alpha^4}{9\alpha^2}\right]+\frac{\alpha^4}{3}\left(\frac{{\tilde r}^2}{a^2}-3\right)\ln\left(\frac{1+\sqrt{1-\alpha^2}}{\alpha}\right)\right\}^{-1/4}\,.
\end{split}
\end{equation}

\begin{figure}
\centering
\includegraphics[scale = 0.45]{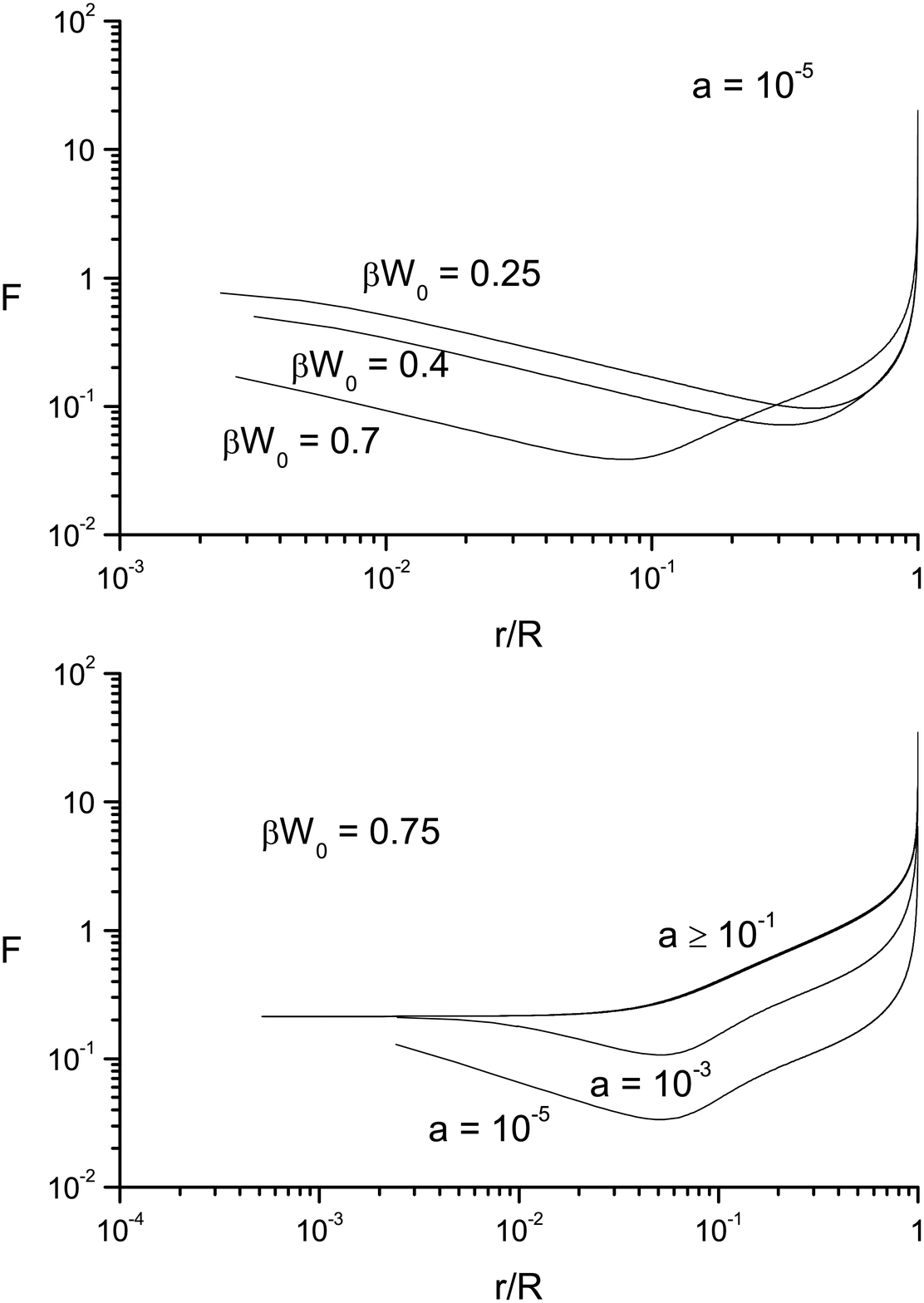}
\caption{Behavior of $F(\alpha_0,a,\tilde r)$ as a function of the relative radius $r/R$, in the limit of full degeneracy and in accordance with Eq.(\ref{F_abeta}). Top panel: $a = 10^{-5}$ for different values of $\beta W_0$. Bottom panel: $\beta W_0 =0.75$ for different values of $a$.}
\label{fig:F_profile}
\end{figure}

Figures \ref{fig:mass_particles} and \ref{fig:F_profile} show the results obtained by Eq.(\ref{massa_3}) and the behavior of the function $F$, which is depending on $\beta W_0$, $a$ and $\tilde r$. This point becomes clear by considering the expression of the density $\tilde\rho$ through the integral $I_{\rho k}$ in the limit of full degeneracy, where $g(x,W)\rightarrow 1$. In fact the thermodynamic quantities in the dimensionless form (\ref{n_less})-(\ref{Pt_less}) and the equations of gravitational equilibrium (\ref{equilibrium_less}) become a function of the product $\beta W$ (or $\alpha$), which is the only parameter changing along the configuration. Moreover, by maintaining the most general character of the treatment, being interested to systems of fermions in GR, we can see that the range of obtained particle masses could describe and limit very massive candidates for the dark matter like the neutralino or other ones (for more details see \cite{2010bertone_book, feng2010}). The very high values of the masses at larger densities shown in Tables \ref{tab:table4} and \ref{tab:table5} indicate that the relativistic description proposed in this paper is necessary to describe systems at so high energetic scales. Other proposals concerning dark matter particles, like sterile neutrinos (de Vega and Sanchez, \cite{deV_sanchez}), can be found by using our model in the Newtonian limit $\beta W \rightarrow 0$ (see also Paper I). Really, from Table \ref{tab:table4} (Newtonian regime and large anisotropy) and for densities $\rho_h < 10^{-6}$ g/cm$^3$, it is possible to obtain lower limits including also masses as $m \sim 10^{-6}$ GeV, namely of the same order of magnitude of the mass value proposed by de Vega and Sanchez.

\section{CONCLUSIONS} \label{conclusions}
\begin{figure}
\centering
\includegraphics[scale = 0.45]{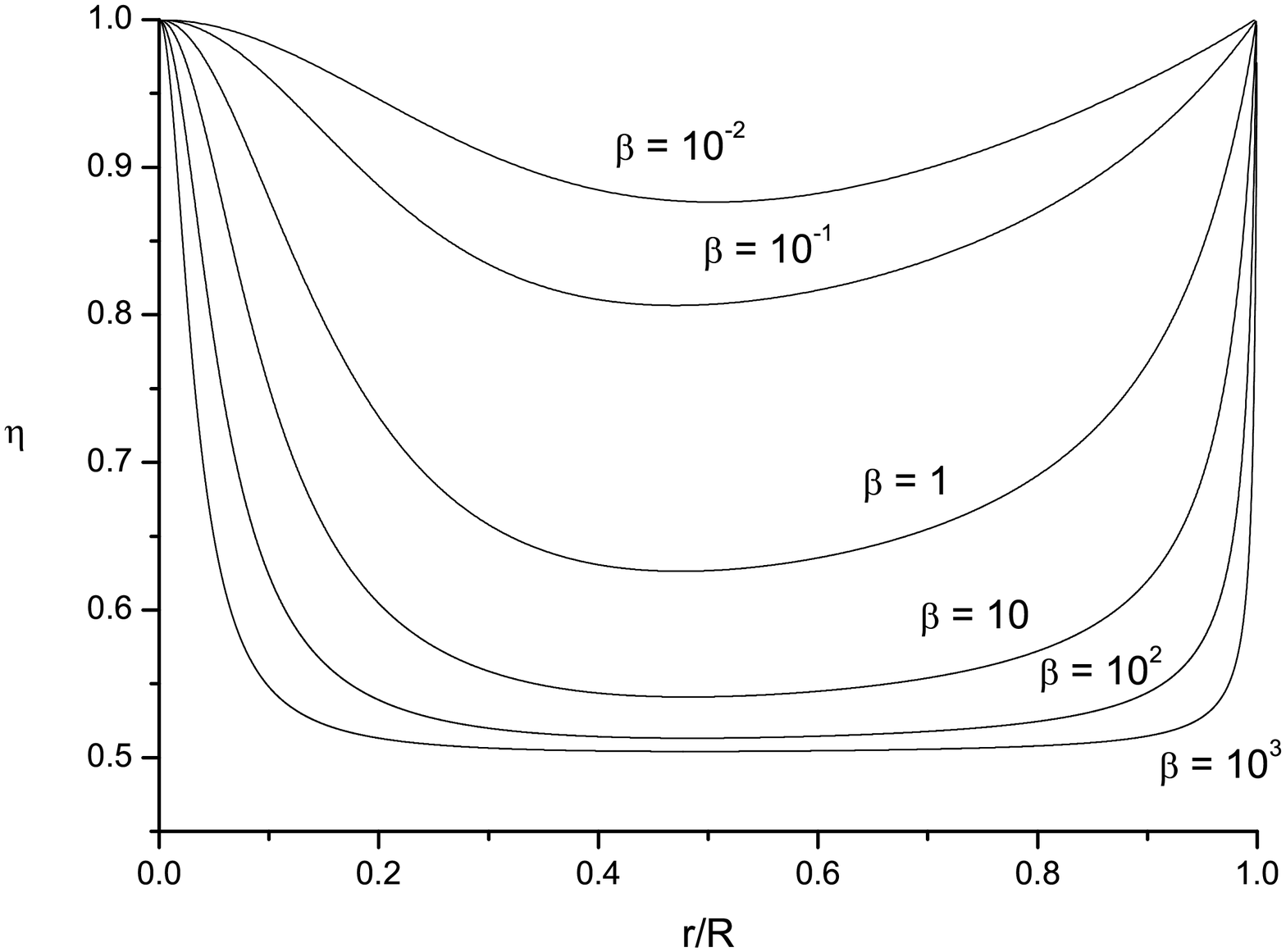}
\caption{Values of the ratio of the velocities $\eta$ for different values of $\beta$, with \emph{a} = $10^{-1}$ and $\beta W_0 =0.3$.}
\label{fig:eta_beta}
\end{figure}

The goal of this paper is to provide a complete understanding of the properties of the systems composed by anisotropic self-gravitating fermions in GR regime. To make an analysis of the various aspects, we considered the effects of the anisotropy on the motion of the particles through the parameter $\eta$ and on the distribution of the matter through the density profiles and the mass versus central density diagrams. In addition, we have classified the configurations through the main parameters by establishing in Sections \ref{subresult_1}, \ref{subresult_2} and \ref{subresult_3} a ``ranking of their influence" in accordance with the order $\beta$, \emph{a} and $\theta_0$. By referring to the motion of the particles, we have recovered the same behavior found in the Paper I and, in the classical limit, the same results obtained in BKMV10. In particular, Fig. \ref{fig:eta_beta} shows the importance of the influence of the parameter $\beta$ on the anisotropy, emphasizing that configurations with an higher ``degree of relativity", precisely described by the parameter $\beta$, present a strong prevalence of tangential motion.

\begin{figure}
\centering
\includegraphics[scale = 0.45]{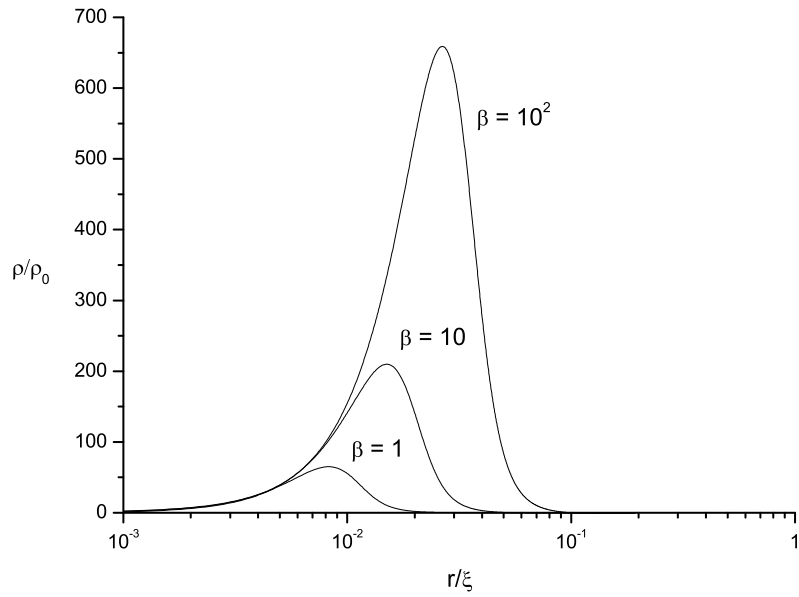}
\caption{Relative density $\rho/\rho_{0}$ for values of $\beta\geq 1$, with \emph{a} = $10^{-3}$ and $\beta W_{0} = 0.6$.}
\label{fig:profile2}
\end{figure}

\begin{figure}
\centering
\includegraphics[scale = 0.45]{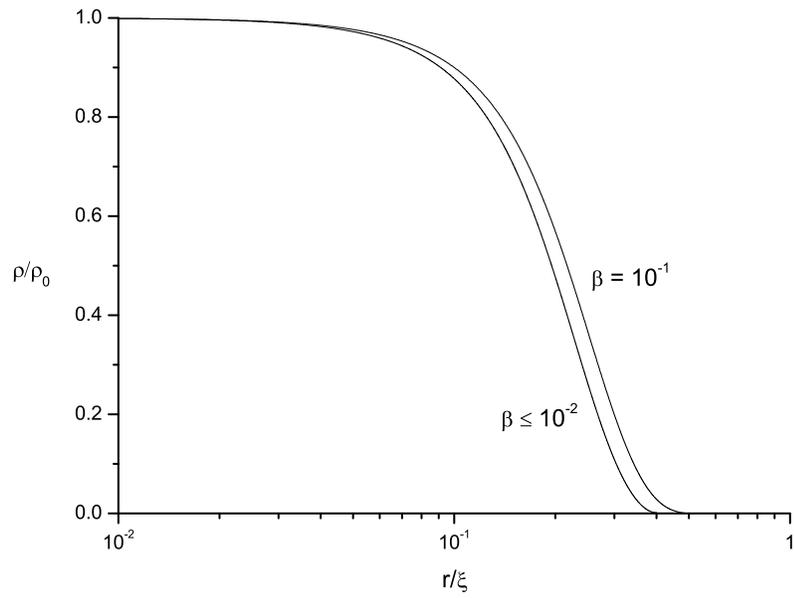}
\caption{The same as in Fig. \ref{fig:profile2} at small values of $\beta$, for \emph{a} = $10^{-1}$ and $\beta W_{0} = 0.1$.}
\label{fig:profile3}
\end{figure}

If we now consider the density profiles, we see clearly the behavior typical of the hollow systems \cite{nguyen_ped13, 1991ralsmith} and, in the Paper I, we have argued the possibility to set the value $a = 0.1$ as the critical threshold for the triggering of the hollowness. However, in the relativistic regime, the things appear quite different, due to the presence of $\beta$. If we look at Fig. \ref{fig:profile3}, we do not see evidences of hollowness; on the contrary, the trend of the profile is very similar to one obtained by Ruffini and Stella \cite{1983rufs}, by using the distribution function (\ref{f_m_a}) in the isotropic limit $L_c \rightarrow \infty$. On the contrary, by looking at Fig. \ref{fig:profile2}, we can say that the hollowness is clearly visible at $a=10^{-3}$ with $\beta\geq 1$.

\begin{figure}
\centering
\includegraphics[scale = 0.45]{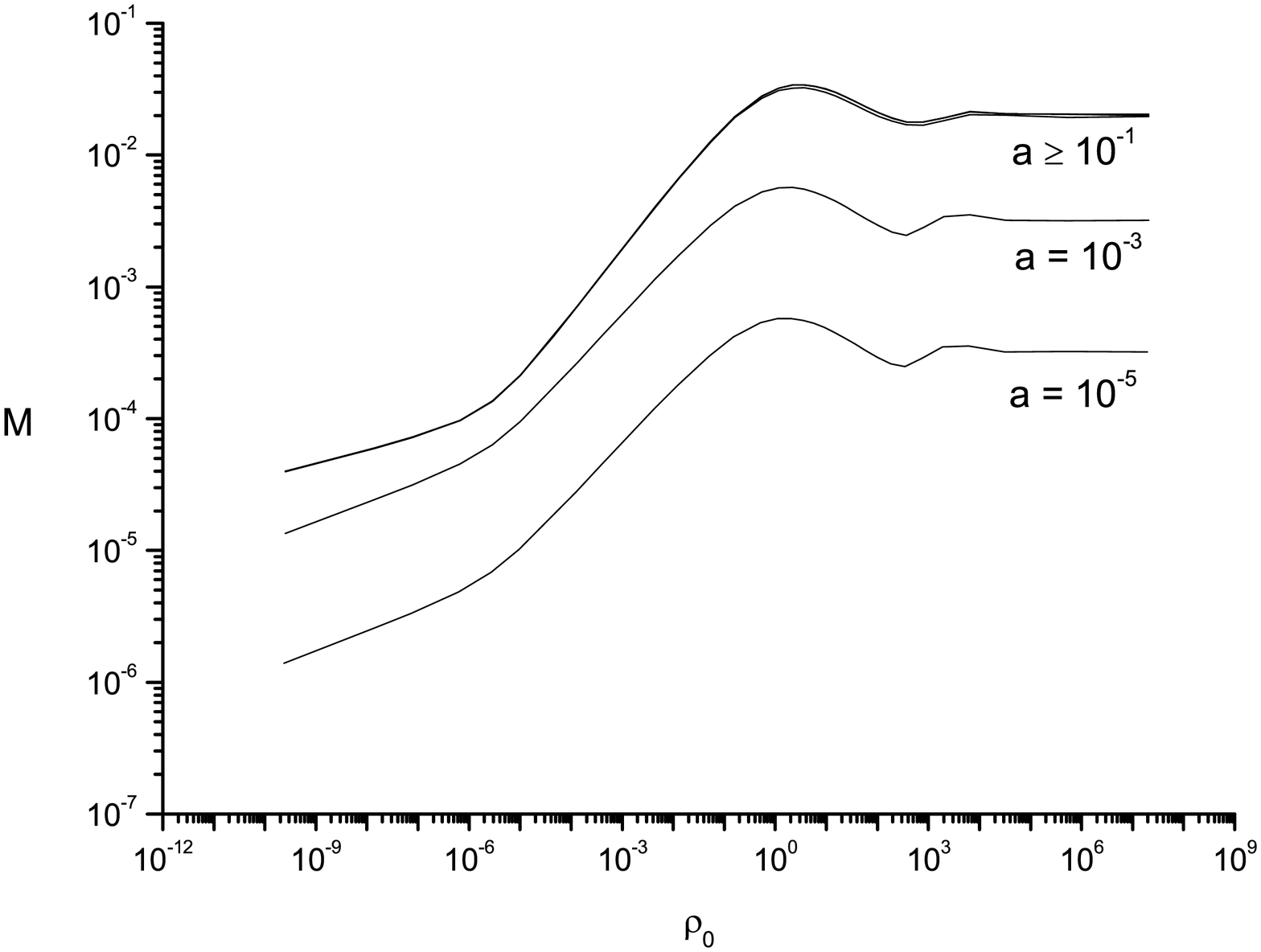}
\caption{Mass of the configurations as a function of the central density at different values of \emph{a}, for $\beta = 10^{-5}$. The quantities are dimensionless.}
\label{fig:M_rho1}
\end{figure}

\begin{figure}
\centering
\includegraphics[scale = 0.45]{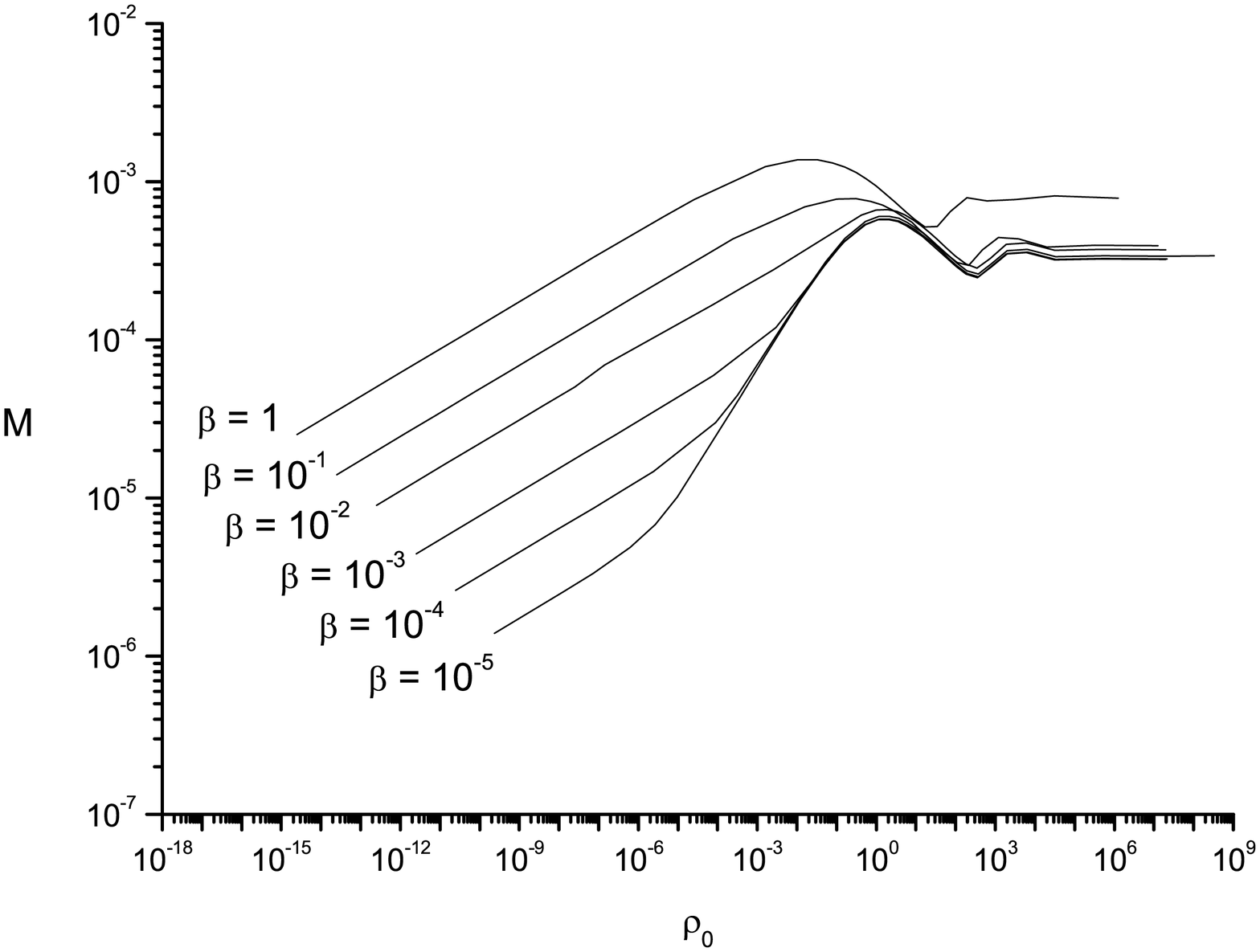}
\caption{Mass of the configurations as a function of the central density at different values of $\beta$, for $a = 10^{-5}$. The quantities are dimensionless.}
\label{fig:M_rho2}
\end{figure}

The more interesting situations are obtained by considering the mass - central density diagrams. In the Fig. \ref{fig:M_rho1}, we can observe how a larger degree of anisotropy reduces the values of the masses of the equilibrium configurations, as well we can further note how, the curve with $a \geq 0.1$ approaches the typical curve of the isotropic systems, indicating, in the relativistic regime, that $a = 0.1$ can be viewed as a good approximation of the isotropic limit for the equilibrium configurations. If we now change the incidence of $\beta$ (see Fig. \ref{fig:M_rho2}), we can note that the configurations tend to a universal curve for large value of the central density while, for decreasing central densities, the degeneracy appears and the curves split, according to the degeneracy level. We have also to mention that, for small values of the central density, the configurations become more and more classical, by gradually leaving their quantum behavior.

\begin{figure}
\centering
\includegraphics[scale = 0.45]{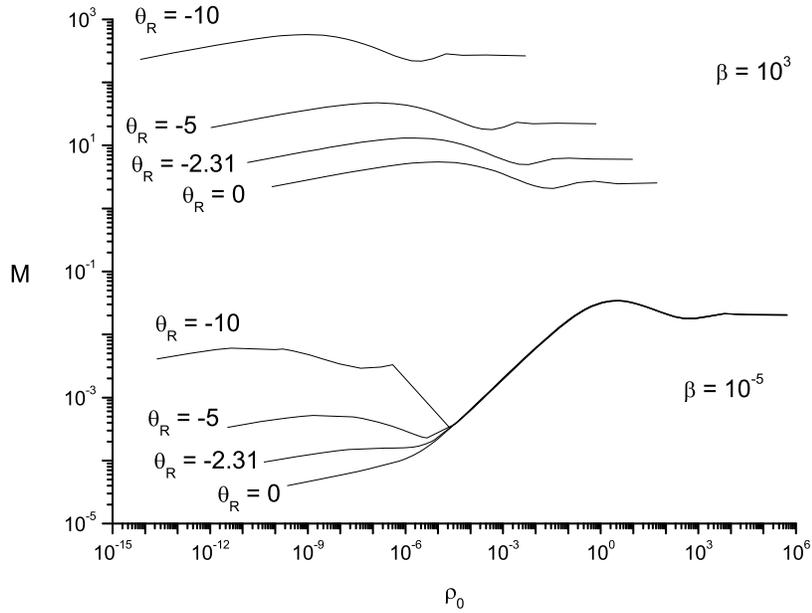}
\caption{The same as in Fig. \ref{fig:M_rho2}, at different values of $\theta_R$ and in the isotropic limit $a\rightarrow\infty$.}
\label{fig:M_rho3}
\end{figure}

Nevertheless, a representation of the influence of the degeneracy level is shown in Fig. \ref{fig:M_rho3}, where we have chosen different values of $\theta_R$, in order to show the gradual passage from the degenerate configurations to the semidegenerate and the classical ones. We have also chosen the two extremal values of $\beta$, representing the Newtonian ($\beta = 10^{-5}$) and the full relativistic ($\beta = 10^3$) limit. Newtonian configurations split up at $\tilde{\rho}_0 \sim 10^{-5}$, showing different behaviors. For $\theta_R <-2.31$, the curves exhibit a minimum that is missing for $\theta_R\geq -2.31$ and, furthermore, the curve with $\theta_R = -10$, corresponding to classical configurations, shows, before the point of minimum, the presence of two local maxima. If we look at the fully relativistic configurations, we see, instead, a regular behavior of the curves, indicating the effects of relativity are much stronger than ones due to the degeneracy level.

\begin{figure}
\centering
\includegraphics[scale = 0.45]{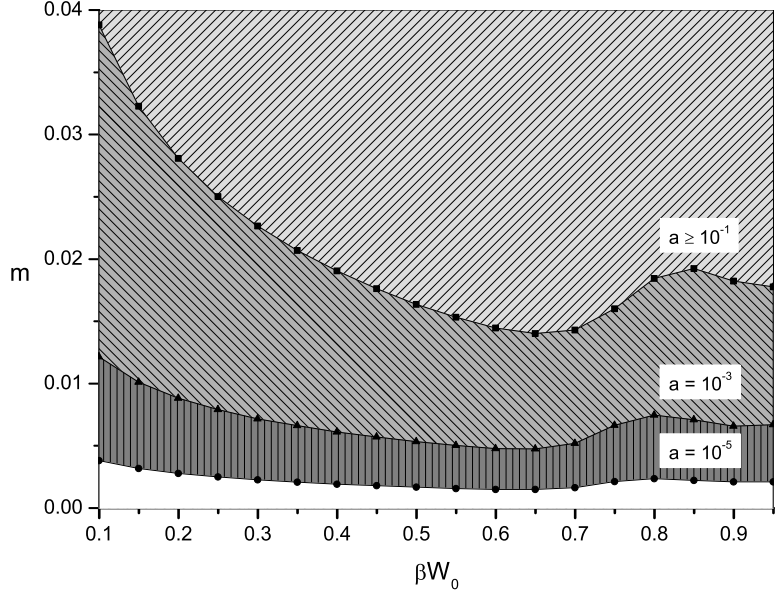}
\caption{Lower limit on mass of the particles $m$, in the limit of full degeneracy and in accordance with Eq.(\ref{massa_3}), as a function of $\beta W_0$ for different values of the parameter $a$. The chosen density, corresponding to the half mass radius, is $\rho_h = 10^{9}$ g/cm$^3$. The masses are expressed in GeV. The different regions characterize the range of available values of the particle mass. A larger anisotropy admits a wider region of availability towards lower mass values.}
\label{fig:mass_particles}
\end{figure}

Moreover, it is important to emphasize that in the limit of full degeneracy we have found an expression that relates the anisotropy with the lower limit on the mass $m$ of the particles [see Eq.(\ref{massa_3})], involving values in the GeV scale. This behavior is clarified in Fig. \ref{fig:mass_particles} where the lower limit of $m$ is plotted as a function of the product $\beta W_0$ for different degrees of anisotropy. We observe an initial decrease of $m$ for $0.1 < \beta W_0 < 0.6$ and then an increase until the achievement of a maximum for $\beta W_0 \sim 0.8$. 

Furthermore, the values we carried out in the Newtonian limit are in good agreement with the results found by de Vega \emph{et al.} \cite{deV_sal_sanchez}, where a serious and detailed study about the particle mass of the dark matter had been performed by also distinguishing the three different case of the Bose-Einstein, Fermi-Dirac and Maxwell-Boltzmann statistics. The fact that the dark matter is composed by particles in the keV scale is a further confirmation that Newtonian gravity could provide a better description for this kind of systems.

Finally, it is also necessary to study the dynamic stability of the equilibrium configurations constructed in this paper and in the Paper I. In spite of the presence of the anisotropy in Eq.(\ref{f_m_a}), we could draw conclusions about the dynamical stability by advancing a general criterion for the anisotropic systems in terms of the polytropic exponent $\gamma$. This problem will be addressed to a forthcoming publication.

\vfill\eject

\bibliography{biblio}
\vfill\eject

\begin{table}
\caption{Some numerical characteristics of fermions for $a = 10^{-1}$ and different values of $\beta W_{0}$, $\beta$, $W_{0}$ and $\theta_{0}$. $\tilde{R}$ and $\tilde{M}$ are the dimensionless radius and mass of the equilibrium configurations, respectively.}
\label{tab:table1}
\begin{center}
\begin{tabular}{>{\centering\arraybackslash}p{2.2cm}>{\centering\arraybackslash}p{2.2cm}>{\centering\arraybackslash}p{2.2cm}>{\centering\arraybackslash}p{2.2cm}>{\centering\arraybackslash}p{2.2cm}>{\centering\arraybackslash}p{2.2cm}}
\hline
\hline
$\beta W_{0}$ & $\beta$ & $W_{0}$ & $\theta_{0}$ & $\tilde{R}$ & $\tilde{M}$ \\ 
\hline
0.1 & $10^{-5}$ & $10^{4}$ & $10^{4}$ & 4.13 $\times 10^{-1}$ & 2.73 $\times 10^{-2}$ \\ 
 & $10^{-4}$ & $10^{3}$ & $10^{3}$ & 4.14 $\times 10^{-1}$ & 2.73 $\times 10^{-2}$ \\ 
 & $10^{-3}$ & $10^{2}$ & $10^{2}$ & 4.29 $\times 10^{-1}$ & 2.72 $\times 10^{-2}$ \\ 
 & $10^{-2}$ & 10 & 10 & 5.22 $\times 10^{-1}$ & 2.75 $\times 10^{-2}$ \\ 
 & $10^{-1}$ & 1 & 1 & 1.09 $\times 10^{0}$ & 5.09 $\times 10^{-2}$ \\ 
 & 1 & $10^{-1}$ & $10^{-1}$ & 2.01 $\times 10^{0}$ & 1.17 $\times 10^{-1}$ \\ 
 & 10 & $10^{-2}$ & $10^{-2}$ & 3.56 $\times 10^{0}$ & 2.31 $\times 10^{-1}$ \\ 
 & $10^{2}$ & $10^{-3}$ & 0 & 6.31 $\times 10^{0}$ & 4.24 $\times 10^{-1}$ \\ 
 & $10^{3}$ & $10^{-4}$ & 0 & 1.12 $\times 10^{1}$ & 7.62 $\times 10^{-1}$ \\ 
0.2 & $10^{-5}$ & 2 $\times 10^{4}$ & 2 $\times 10^{4}$ & 2.94 $\times 10^{-1}$ & 3.25 $\times 10^{-2}$ \\ 
 & $10^{-4}$ & 2 $\times 10^{3}$ & 2 $\times 10^{3}$ & 2.94 $\times 10^{-1}$ & 3.25 $\times 10^{-2}$ \\ 
 & $10^{-3}$ & 2 $\times 10^{2}$ & 2 $\times 10^{2}$ & 2.97 $\times 10^{-1}$ & 3.25 $\times 10^{-2}$ \\ 
 & $10^{-2}$ & 2 $\times 10^{1}$ & 2 $\times 10^{1}$ & 3.34 $\times 10^{-1}$ & 3.29 $\times 10^{-2}$ \\ 
 & $10^{-1}$ & 2 $\times 10^{0}$ & 2 $\times 10^{0}$ & 6.27 $\times 10^{-1}$ & 4.87 $\times 10^{-2}$ \\ 
 & 1 & 2 $\times 10^{-1}$ & 2 $\times 10^{-1}$ & 1.19 $\times 10^{0}$ & 1.11 $\times 10^{-1}$ \\ 
 & 10 & 2 $\times 10^{-2}$ & 0 & 2.09 $\times 10^{0}$ & 2.21 $\times 10^{-1}$ \\ 
 & $10^{2}$ & 2 $\times 10^{-3}$ & 0 & 3.68 $\times 10^{0}$ & 4.06 $\times 10^{-1}$ \\ 
 & $10^{3}$ & 2 $\times 10^{-4}$ & 0 & 6.53 $\times 10^{0}$ & 7.30 $\times 10^{-1}$ \\ 
0.3 & $10^{-5}$ & 3 $\times 10^{4}$ & 3 $\times 10^{4}$ & 2.29 $\times 10^{-1}$ & 3.16 $\times 10^{-2}$ \\ 
 & $10^{-4}$ & 3 $\times 10^{3}$ & 3 $\times 10^{3}$ & 2.29 $\times 10^{-1}$ & 3.16 $\times 10^{-2}$ \\ 
 & $10^{-3}$ & 3 $\times 10^{2}$ & 3 $\times 10^{2}$ & 2.31 $\times 10^{-1}$ & 3.17 $\times 10^{-2}$ \\ 
 & $10^{-2}$ & 3 $\times 10^{1}$ & 3 $\times 10^{1}$ & 2.52 $\times 10^{-1}$ & 3.21 $\times 10^{-2}$ \\ 
 & $10^{-1}$ & 3 $\times 10^{0}$ & 3 $\times 10^{0}$ & 4.41 $\times 10^{-1}$ & 4.31 $\times 10^{-2}$ \\ 
 & 1 & 3 $\times 10^{-1}$ & 3 $\times 10^{-1}$ & 8.54 $\times 10^{-1}$ & 9.64 $\times 10^{-2}$ \\ 
 & 10 & 3 $\times 10^{-2}$ & 0 & 1.49 $\times 10^{0}$ & 1.91 $\times 10^{-1}$ \\ 
 & $10^{2}$ & 3 $\times 10^{-3}$ & 0 & 2.61 $\times 10^{0}$ & 3.51 $\times 10^{-1}$ \\ 
 & $10^{3}$ & 3 $\times 10^{-4}$ & 0 & 4.62 $\times 10^{0}$ & 6.30 $\times 10^{-1}$ \\
\hline
\end{tabular}
\end{center}
\end{table}

\begin{table}
\caption{The same as Tab. \ref{tab:table1}, for $a = 10^{-3}$.}
\label{tab:table2}
\begin{center}
\begin{tabular}{>{\centering\arraybackslash}p{2.2cm}>{\centering\arraybackslash}p{2.2cm}>{\centering\arraybackslash}p{2.2cm}>{\centering\arraybackslash}p{2.2cm}>{\centering\arraybackslash}p{2.2cm}>{\centering\arraybackslash}p{2.2cm}}
\hline
\hline
$\beta W_{0}$ & $\beta$ & $W_{0}$ & $\theta_{0}$ & $\tilde{R}$ & $\tilde{M}$ \\ 
\hline
0.4 & $10^{-5}$ & 4 $\times 10^{4}$ & 4 $\times 10^{4}$ & 2.39 $\times 10^{-2}$ & 4.50 $\times 10^{-3}$ \\ 
 & $10^{-4}$ & 4 $\times 10^{3}$ & 4 $\times 10^{3}$ & 2.39 $\times 10^{-2}$ & 4.50 $\times 10^{-3}$ \\ 
 & $10^{-3}$ & 4 $\times 10^{2}$ & 4 $\times 10^{2}$ & 2.41 $\times 10^{-2}$ & 4.51 $\times 10^{-3}$ \\ 
 & $10^{-2}$ & 4 $\times 10^{1}$ & 4 $\times 10^{1}$ & 2.54 $\times 10^{-2}$ & 4.58 $\times 10^{-3}$ \\ 
 & $10^{-1}$ & 4 $\times 10^{0}$ & 4 $\times 10^{0}$ & 3.60 $\times 10^{-2}$ & 5.46 $\times 10^{-3}$ \\ 
 & 1 & 4 $\times 10^{-1}$ & 4 $\times 10^{-1}$ & 6.42 $\times 10^{-2}$ & 9.26 $\times 10^{-3}$ \\ 
 & 10 & 4 $\times 10^{-2}$ & 0 & 1.15 $\times 10^{-1}$ & 1.66 $\times 10^{-2}$ \\ 
 & $10^{2}$ & 4 $\times 10^{-3}$ & 0 & 2.03 $\times 10^{-1}$ & 2.93 $\times 10^{-2}$ \\ 
 & $10^{3}$ & 4 $\times 10^{-4}$ & 0 & 3.60 $\times 10^{-1}$ & 5.21 $\times 10^{-2}$ \\ 
0.5 & $10^{-5}$ & 5 $\times 10^{4}$ & 5 $\times 10^{4}$ & 2.07 $\times 10^{-2}$ & 3.67 $\times 10^{-3}$ \\ 
 & $10^{-4}$ & 5 $\times 10^{3}$ & 5 $\times 10^{3}$ & 2.07 $\times 10^{-2}$ & 3.69 $\times 10^{-3}$ \\ 
 & $10^{-3}$ & 5 $\times 10^{2}$ & 5 $\times 10^{2}$ & 2.08 $\times 10^{-2}$ & 3.69 $\times 10^{-3}$ \\ 
 & $10^{-2}$ & 5 $\times 10^{1}$ & 5 $\times 10^{1}$ & 2.19 $\times 10^{-2}$ & 3.75 $\times 10^{-3}$ \\ 
 & $10^{-1}$ & 5 $\times 10^{0}$ & 5 $\times 10^{0}$ & 3.09 $\times 10^{-2}$ & 4.40 $\times 10^{-3}$ \\ 
 & 1 & 5 $\times 10^{-1}$ & 5 $\times 10^{-1}$ & 5.57 $\times 10^{-2}$ & 7.37 $\times 10^{-3}$ \\ 
 & 10 & 5 $\times 10^{-2}$ & 0 & 9.97 $\times 10^{-2}$ & 1.32 $\times 10^{-2}$ \\ 
 & $10^{2}$ & 5 $\times 10^{-3}$ & 0 & 1.76 $\times 10^{-1}$ & 2.33 $\times 10^{-2}$ \\ 
 & $10^{3}$ & 5 $\times 10^{-4}$ & 0 & 3.13 $\times 10^{-1}$ & 4.14 $\times 10^{-2}$ \\ 
0.6 & $10^{-5}$ & 6 $\times 10^{4}$ & 6 $\times 10^{4}$ & 2.04 $\times 10^{-2}$ & 2.92 $\times 10^{-3}$ \\ 
 & $10^{-4}$ & 6 $\times 10^{3}$ & 6 $\times 10^{3}$ & 2.04 $\times 10^{-2}$ & 2.92 $\times 10^{-3}$ \\ 
 & $10^{-3}$ & 6 $\times 10^{2}$ & 6 $\times 10^{2}$ & 2.05 $\times 10^{-2}$ & 2.93 $\times 10^{-3}$ \\ 
 & $10^{-2}$ & 6 $\times 10^{1}$ & 6 $\times 10^{1}$ & 2.17 $\times 10^{-2}$ & 2.96 $\times 10^{-3}$ \\ 
 & $10^{-1}$ & 6 $\times 10^{0}$ & 6 $\times 10^{0}$ & 3.22 $\times 10^{-2}$ & 3.44 $\times 10^{-3}$ \\ 
 & 1 & 6 $\times 10^{-1}$ & 6 $\times 10^{-1}$ & 6.05 $\times 10^{-2}$ & 5.73 $\times 10^{-3}$ \\ 
 & 10 & 6 $\times 10^{-2}$ & 0 & 1.09 $\times 10^{-1}$ & 1.03 $\times 10^{-2}$ \\ 
 & $10^{2}$ & 6 $\times 10^{-3}$ & 0 & 1.92 $\times 10^{-1}$ & 1.81 $\times 10^{-2}$ \\ 
 & $10^{3}$ & 6 $\times 10^{-4}$ & 0 & 3.40 $\times 10^{-1}$ & 3.22 $\times 10^{-2}$ \\
\hline
\end{tabular}
\end{center}
\end{table}

\begin{table}
\caption{The same as Tab. \ref{tab:table1}, for $a = 10^{-5}$.}
\label{tab:table3}
\begin{center}
\begin{tabular}{>{\centering\arraybackslash}p{2.2cm}>{\centering\arraybackslash}p{2.2cm}>{\centering\arraybackslash}p{2.2cm}>{\centering\arraybackslash}p{2.2cm}>{\centering\arraybackslash}p{2.2cm}>{\centering\arraybackslash}p{2.2cm}}
\hline
\hline
$\beta W_{0}$ & $\beta$ & $W_{0}$ & $\theta_{0}$ & $\tilde{R}$ & $\tilde{M}$ \\ 
\hline
0.7 & $10^{-5}$ & 7 $\times 10^{4}$ & 7 $\times 10^{4}$ & 2.95 $\times 10^{-3}$ & 2.48 $\times 10^{-4}$ \\
 & $10^{-4}$ & 7 $\times 10^{3}$ & 7 $\times 10^{3}$ & 2.96 $\times 10^{-3}$ & 2.48 $\times 10^{-4}$ \\ 
 & $10^{-3}$ & 7 $\times 10^{2}$ & 7 $\times 10^{2}$ & 2.99 $\times 10^{-3}$ & 2.48 $\times 10^{-4}$ \\ 
 & $10^{-2}$ & 7 $\times 10^{1}$ & 7 $\times 10^{1}$ & 3.29 $\times 10^{-3}$ & 2.49 $\times 10^{-4}$ \\ 
 & $10^{-1}$ & 7 $\times 10^{0}$ & 7 $\times 10^{0}$ & 6.38 $\times 10^{-3}$ & 2.94 $\times 10^{-4}$ \\ 
 & 1 & 7 $\times 10^{-1}$ & 7 $\times 10^{-1}$ & 1.38 $\times 10^{-2}$ & 5.13 $\times 10^{-4}$ \\ 
 & 10 & 7 $\times 10^{-2}$ & 0 & 2.49 $\times 10^{-2}$ & 9.19 $\times 10^{-4}$ \\ 
 & $10^{2}$ & 7 $\times 10^{-3}$ & 0 & 4.38 $\times 10^{-2}$ & 1.62 $\times 10^{-3}$ \\ 
 & $10^{3}$ & 7 $\times 10^{-4}$ & 0 & 7.78 $\times 10^{-1}$ & 2.89 $\times 10^{-3}$ \\ 
0.8 & $10^{-5}$ & 8 $\times 10^{4}$ & 8 $\times 10^{4}$ & 3.52 $\times 10^{-3}$ & 3.49 $\times 10^{-4}$ \\ 
 & $10^{-4}$ & 8 $\times 10^{3}$ & 8 $\times 10^{3}$ & 3.53 $\times 10^{-3}$ & 3.49 $\times 10^{-4}$ \\ 
 & $10^{-3}$ & 8 $\times 10^{2}$ & 8 $\times 10^{2}$ & 3.56 $\times 10^{-3}$ & 3.49 $\times 10^{-4}$ \\ 
 & $10^{-2}$ & 8 $\times 10^{1}$ & 8 $\times 10^{1}$ & 3.96 $\times 10^{-3}$ & 3.51 $\times 10^{-4}$ \\ 
 & $10^{-1}$ & 8 $\times 10^{0}$ & 8 $\times 10^{0}$ & 6.86 $\times 10^{-3}$ & 4.41 $\times 10^{-4}$ \\ 
 & 1 & 8 $\times 10^{-1}$ & 8 $\times 10^{-1}$ & 1.24 $\times 10^{-2}$ & 7.81 $\times 10^{-4}$ \\ 
 & 10 & 8 $\times 10^{-2}$ & 0 & 2.25 $\times 10^{-2}$ & 1.41 $\times 10^{-3}$ \\ 
 & $10^{2}$ & 8 $\times 10^{-3}$ & 0 & 3.92 $\times 10^{-2}$ & 2.48 $\times 10^{-3}$ \\ 
 & $10^{3}$ & 8 $\times 10^{-4}$ & 0 & 6.98 $\times 10^{-2}$ & 4.39 $\times 10^{-3}$ \\ 
0.9 & $10^{-5}$ & 9 $\times 10^{4}$ & 9 $\times 10^{4}$ & 2.85 $\times 10^{-3}$ & 3.21 $\times 10^{-4}$ \\ 
 & $10^{-4}$ & 9 $\times 10^{3}$ & 9 $\times 10^{3}$ & 2.85 $\times 10^{-3}$ & 3.21 $\times 10^{-4}$ \\ 
 & $10^{-3}$ & 9 $\times 10^{2}$ & 9 $\times 10^{2}$ & 2.88 $\times 10^{-3}$ & 3.21 $\times 10^{-4}$ \\ 
 & $10^{-2}$ & 9 $\times 10^{1}$ & 9 $\times 10^{1}$ & 3.12 $\times 10^{-3}$ & 3.24 $\times 10^{-4}$ \\ 
 & $10^{-1}$ & 9 $\times 10^{0}$ & 9 $\times 10^{0}$ & 5.23 $\times 10^{-3}$ & 3.84 $\times 10^{-4}$ \\ 
 & 1 & 9 $\times 10^{-1}$ & 9 $\times 10^{-1}$ & 1.00 $\times 10^{-2}$ & 6.59 $\times 10^{-4}$ \\ 
 & 10 & 9 $\times 10^{-2}$ & 0 & 1.81 $\times 10^{-2}$ & 1.19 $\times 10^{-3}$ \\ 
 & $10^{2}$ & 9 $\times 10^{-3}$ & 0 & 3.16 $\times 10^{-2}$ & 2.08 $\times 10^{-3}$ \\ 
 & $10^{3}$ & 9 $\times 10^{-4}$ & 0 & 5.65 $\times 10^{-2}$ & 3.71 $\times 10^{-3}$ \\
\hline
\end{tabular}
\end{center}
\end{table}

\begin{table}
\caption{Lower limits on the mass of particles by Eq.(\ref{massa_3}) (full degeneracy) for $a=10^{-5}$ (large anisotropy) and $\beta W_0 = 10^{-6}$ (Newtonian regime). The densities $\rho_h$ are expressed in g/cm$^3$, the masses $m$ in GeV.}
\label{tab:table4}
\begin{center}
\begin{tabular}{>{\centering\arraybackslash}p{2.5cm}>{\centering\arraybackslash}p{2.5cm}>{\centering\arraybackslash}p{2.5cm}>{\centering\arraybackslash}p{2.5cm}}
\hline
\hline
$\rho_h$ & \emph{m} & $\rho_h$ & \emph{m} \\ 
\hline 
$10^{-21}$ & 9.20 $\times$ $10^{-10}$ & $10^{6}$ & 5.17 $\times$ $10^{-3}$ \\ 
$10^{-18}$ & 5.17 $\times$ $10^{-9}$ & $10^{9}$ & 2.91 $\times$ $10^{-2}$ \\ 
$10^{-15}$ & 2.91 $\times$ $10^{-8}$ & $10^{12}$ & 1.64 $\times$ $10^{-1}$ \\ 
$10^{-12}$ & 1.64 $\times$ $10^{-7}$ & $10^{15}$ & 9.20 $\times$ $10^{-1}$ \\ 
$10^{-9}$ & 9.20 $\times$ $10^{-7}$ & $10^{18}$ & 5.17 $\times$ $10^{0}$ \\ 
$10^{-6}$ & 5.17 $\times$ $10^{-6}$ & $10^{21}$ & 2.91 $\times$ $10^{1}$ \\ 
$10^{-3}$ & 2.91 $\times$ $10^{-5}$ & $10^{24}$ & 1.64 $\times$ $10^{2}$ \\ 
$10^{0}$ & 1.64 $\times$ $10^{-4}$ & $10^{27}$ & 9.20 $\times$ $10^{2}$ \\ 
$10^{3}$ & 9.20 $\times$ $10^{-4}$ & $10^{30}$ & 5.17 $\times$ $10^{3}$ \\ 
\hline
\end{tabular}
\end{center}
\end{table}

\begin{table}
\caption{The same as in Table \ref{tab:table4} (full degeneracy and Newtonian regime) for $a\geq 10^{-1}$ (isotropic limit). Also in this case the densities $\rho_h$ are expressed in g/cm$^3$ and the masses $m$ in GeV.}
\label{tab:table5}
\begin{center}
\begin{tabular}{>{\centering\arraybackslash}p{2.5cm}>{\centering\arraybackslash}p{2.5cm}>{\centering\arraybackslash}p{2.5cm}>{\centering\arraybackslash}p{2.5cm}}
\hline
\hline
$\rho_h$ & \emph{m} & $\rho_h$ & \emph{m} \\ 
\hline 
$10^{-21}$ & 7.83 $\times$ $10^{-9}$ & $10^{6}$ & 4.40 $\times$ $10^{-2}$ \\ 
$10^{-18}$ & 4.40 $\times$ $10^{-8}$ & $10^{9}$ & 2.48 $\times$ $10^{-1}$ \\ 
$10^{-15}$ & 2.48 $\times$ $10^{-7}$ & $10^{12}$ & 1.39 $\times$ $10^{0}$ \\ 
$10^{-12}$ & 1.39 $\times$ $10^{-6}$ & $10^{15}$ & 7.83 $\times$ $10^{0}$ \\ 
$10^{-9}$ & 7.83 $\times$ $10^{-6}$ & $10^{18}$ & 4.40 $\times$ $10^{1}$ \\ 
$10^{-6}$ & 4.40 $\times$ $10^{-5}$ & $10^{21}$ & 2.48 $\times$ $10^{2}$ \\ 
$10^{-3}$ & 2.48 $\times$ $10^{-4}$ & $10^{24}$ & 1.39 $\times$ $10^{3}$ \\ 
$10^{0}$ & 1.39 $\times$ $10^{-3}$ & $10^{27}$ & 7.83 $\times$ $10^{3}$ \\ 
$10^{3}$ & 7.83 $\times$ $10^{-3}$ & $10^{30}$ & 4.40 $\times$ $10^{4}$ \\ 
\hline
\end{tabular}
\end{center}
\end{table}

\end{document}